\documentclass[aps,pre,showpacs,twocolumn,groupedaddress]{revtex4-1}

\usepackage{graphicx}
\usepackage{dcolumn}
\usepackage{bm}
\usepackage{enumerate}
\usepackage{amsmath}
\usepackage{footnote}
\usepackage{amssymb}
\usepackage{natbib}

\begin{document}

\title{ \bf Heterogeneity of time delays determines synchronization of coupled oscillators}

\author{Spase Petkoski$^{1,2}$}
\email[]{spase.petkoski@univ-amu.fr}

\author{Andreas Spiegler$^{1}$}

\author{Timoth\'{e}e Proix$^{1}$}

\author{Parham Aram$^{3}$}

\author{Jean-Jacques Temprado$^{2}$}

\author{Viktor K. Jirsa$^{1}$}

\affiliation{$^1$Aix-Marseille Universit\'{e}, Inserm, INS UMR\_S 1106, 13005, Marseille, France}
\affiliation{$^2$Aix-Marseille Universit\'{e}, CNRS, ISM UMR      7287,  13288, Marseille, France}
\affiliation{$^3$Department of Automatic Control and Systems Engineering, University of Sheffield, Sheffield S10 2TN, UK}

\date{\today}

\begin{abstract}

Network couplings of oscillatory large-scale systems, such as the brain, have a space-time structure composed of connection strengths and signal transmission delays. 
We provide a theoretical framework, which allows treating the spatial distribution of time delays
  with regard to synchronization, by decomposing it into patterns and therefore reducing the stability analysis into the tractable problem of a finite set of delay-coupled differential equations.
We analyse delay-structured networks of  phase oscillators  and we find that, depending on the heterogeneity of the delays, the oscillators group in phase-shifted, anti-phase, steady, and non-stationary clusters, and analytically compute their stability boundaries.  
These results find direct application in the study of 
brain oscillations.  

\end{abstract}


\maketitle

\section{Introduction}
\label{sec:Intro}

Time delays due to finite signal transmission are unavoidable in physical, biological and technical systems. They are often considered a nuisance and can be mostly ignored when they are small with regard to the characteristic time scale of the system. In a number of systems though, foremost in the brain, the  delays (10 to 200 ms) are on the same scale as the signal operation (10 to 250 ms) \cite{Buzsaki:04, nunez2006electric} and contribute critically to the system's spatiotemporal organization. 
Rhythms and their synchronization, as one of the key mechanisms of brain function  \cite{fries2005mechanism,varela2001brainweb}, are ubiquitous in the nervous system and are particularly sensitive to delays, because shifts in phasing may easily change the nature of the mutual influences from excitatory to inhibitory and vice versa. 

The spatiotemporal organization of oscillatory networks is often studied via coupled phase oscillators, which arise 
for
weak interactions 
\cite{Roy2011,Kuramoto:84,Izhikevich1998,Ermentrout2009}. 
Phase models
represent a 
simple class of models for interacting nonlinear limit-cycle oscillators that exhibit richness in behaviour while at the same time admit analytic approaches and a direct link to more complex biophysical models. 
For small delays, the delayed interactions between oscillators are reduced to phase shifts \cite{Izhikevich1998, Ermentrout2009, Ton2014}, but they appear inside the state variables \cite{Izhikevich1998, Ermentrout2009} when delays are of the order of 1/coupling-strength. 
Transmission delays become particularly long in large-scale brain models with biologically realistic connectivity \cite{ghosh2008noise, Deco:09}, which have become feasible with the recent advance of non-invasive structural brain imaging \cite{Hagmann2010, Johansen-Berg2009}. 
Together, the connectivity strengths and time delays, define 
the Connectome as the final determinant of the brain network behavior \cite{jirsa2009neural, sanz2015mathematical}. 
In absence of delays, the importance of  couplings' topology for the synchronization of phase oscillators is well understood, both in random  \cite{arenas2008synchronization, Rodrigues2015}, and in networks structured by
natural frequencies and coupling strengths  \cite{Montbrio:04, sheeba2008routes}. 

Many of the phase network models of the brain use explicit delays 
in the state variable 
(see \cite{Ermentrout2009} for review)
and make simplifying assumptions on either 
their distribution
(such as distance dependence \cite{Crook1997}) or the spatial coupling topology (such as rings in one dimension \cite{Bressloff1997}).  
In this letter we will develop a principled approach for decomposing the coupling's 
structure into modes,
which characterize synchronization 
as a function of the spatial distribution of the time delays. 
Phase reduction of weakly delay-coupled oscillators  with long delays in comparison to the coupling strengths $K_{i,j}$ or natural frequencies $\omega_i$ leads to periodic coupling function with explicit heterogeneous time-delays \cite{Izhikevich1998, Ermentrout2009}. 
The general coupling function may lead to an enormous diversity of collective states \cite{Daido1996} and have been shown to be of interest for the brain \cite{Roy2011,Stankovski2016}.
However, for the reasons of analytical tractability much research keeps only an initial portion of its Fourier series, therefore leading to the Kuramoto model (KM) \cite{Kuramoto:84}.
Considering the KM for symmetric, link-dependent delays, $\tau_{i,j}=\tau_{j,i}$, 
 phases $\theta_i$  of each oscillator evolve as
\begin{eqnarray}
\dot{\theta}_i =\omega_i +\frac{1}{N}\sum_{j=1}^N K_{i,j} \sin[ \theta_j(t-\tau_{i,j}) - \theta_i],   i=1\ldots N, \ \ \ \
 \label{eqn:KMdel}
\end{eqnarray}
where 
 $\omega_i$ follow a probability density function (PDF) $g(\omega)$. 
Recent works on the KM study steady synchronization for arbitrary parameters  \cite{iatsenko2013stationary}, glassy  \cite{iatsenko2014glassy} and chimera \cite{omel2008chimera} states, non-isochronicity \cite{Montbrio:11b}, non-autonomicity \cite{Petkoski:12}, 
and 
networks of
spiking neurons  \cite{pazo2014low, laing2014derivation, montbrio2015macroscopic}. 

\section{Model}
\label{sec:Model}

\begin{figure}[tb!]
\centering
\includegraphics{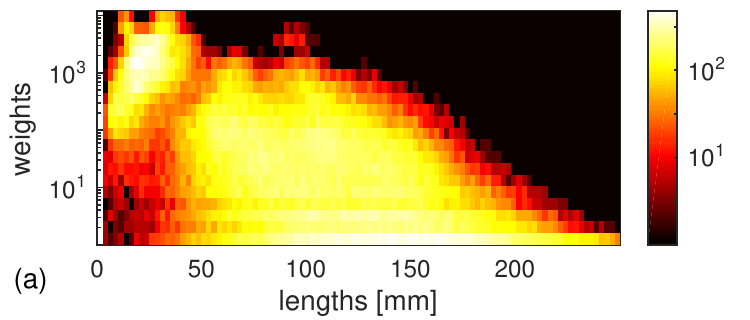} 
\includegraphics{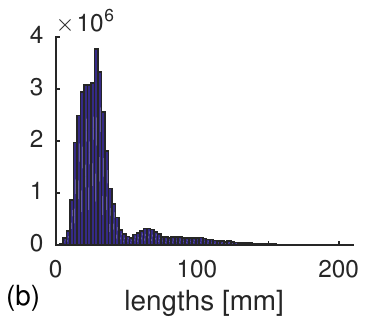} \hspace{1em}
\includegraphics{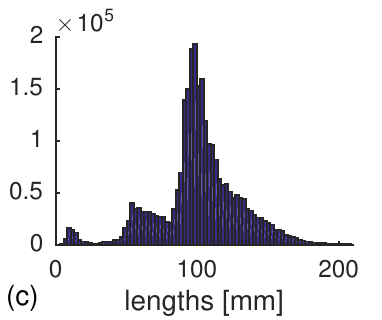} 
\caption{(color online) Tract lengths and weights from 
100 healthy subjects.  
Joint distribution 
(a), and histogram of weighted lengths for (b) intra- and (c) inter-hemisphere links.}
\label{fig:1}
\end{figure}

Upon analysis of human connectomes,  each consisting of  few millions tracts identified with magnetic resonance imaging 
 and connecting 68 cortical regions  \cite{van2013wu}
 where for each link weights are numbers of the individual tracts and lengths are their averages,
the results imply that the lengths of connection routes between brain areas are multimodaly distributed, Fig. \ref{fig:1} (a). 
%
Moreover, the modes in the lengths distribution are spatially heterogeneous and as a first approximation two main modes correspond to the intra- and inter-hemispheric links, Fig. \ref{fig:1} (b, c).
This insight suggests that the complex space-time structure of the connectivity maybe approximated by a less complex mode decomposition in the parameter space, which will aid in the mathematical analysis of the large-scale brain dynamics. 
These ideas have been previously exploited by mean field techniques 
\cite{Assisi2005} in which degrees of freedom are associated with variation of a parameter. 
In neuroscience, this approach has been successfully applied to neural populations with heterogeneous thresholds \cite{Stefanescu2008}. 
We take here an equivalent approach for distributed delays. To emphasize the influence of temporal component of the space-time structure upon the network dynamics, we only consider homogeneous strength for the connections, $K_{i,j}=K$, 
and the extension to distributed connection weights is straight forward. 
Besides, unimodal positive couplings, as observed in Fig. \ref{fig:1} (a), do not bring novel mean-field behavior  for the KM with randomly distributed frequencies  \cite{Paissan:07, petkoski2013mean, arenas2008synchronization}.
The network dynamics is then described by order parameters, which either represent the collective behavior from the instantaneous phases 
\cite{Kuramoto:84}, 
\begin{eqnarray}
 \label{eqn:z}
z(t) \equiv r(t)\mathrm{e}^{i\psi(t)}=N^{-1}\sum_{j} \mathrm{e}^{i\theta_{j}},
\end{eqnarray}
or the delayed, node-dependent mean-field  \cite{lee2009large} that acts as forcing on each oscillator 
\begin{eqnarray}
\xi_i(t)=N^{-1}\sum_{j} \mathrm{e}^{i\theta_j(t-\tau_{i,j})},
 \label{eqn:ksi}
\end{eqnarray}
hereafter referred to as global and local order parameters.

\begin{figure}[tb!]
\centering
\includegraphics[width=0.46\textwidth]{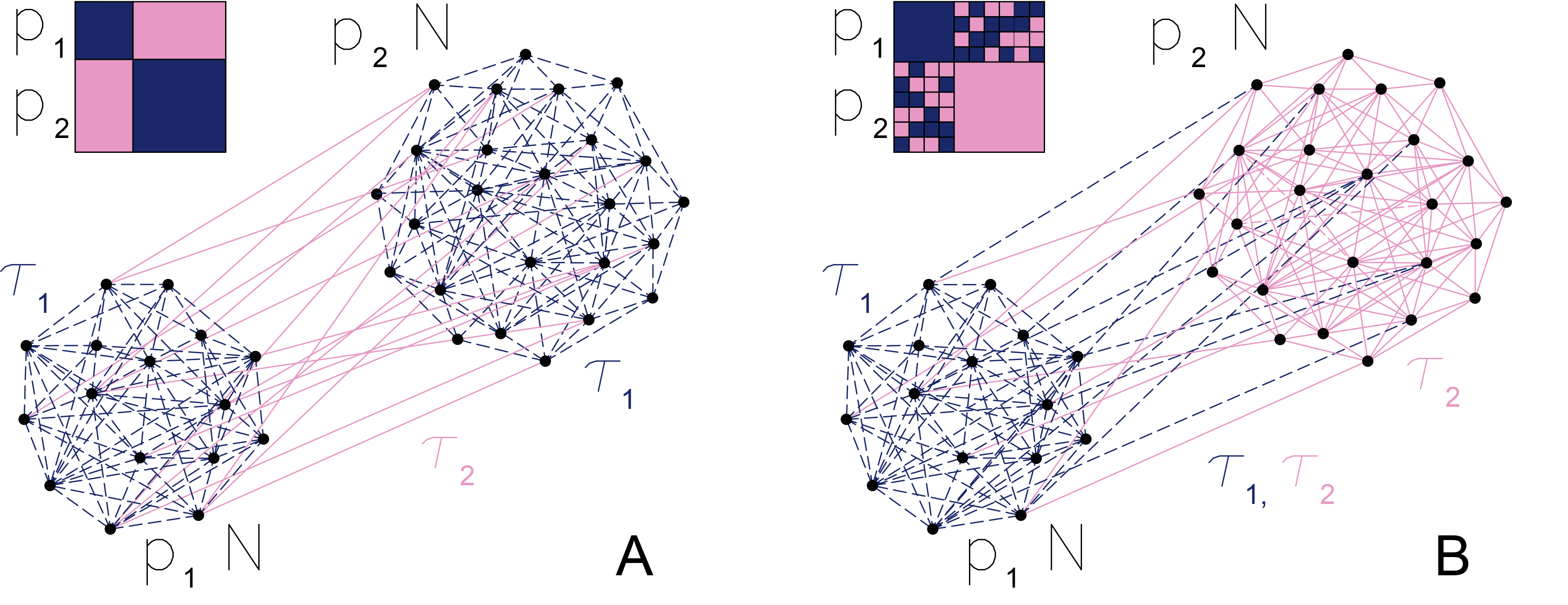} 
\caption{(color online) Sketch of delay-imposed structure and connectivity matrices for oscillators in models A and B with link delays $\tau_1$ dark (blue) dashed lines and $\tau_2$ light (red) lines. 
}
\label{fig:2}
\end{figure}

At least two modes can then be identified from the distribution of tract lengths, Fig. \ref{fig:1} (a), which as first approximation for simplicity is assumed to be bimodal-$\delta$
\begin{eqnarray}
 \label{eqn:h}
 h(\tau)= p_1^{\prime} \delta(\tau-\tau_{1}) + p_2^{\prime} \delta(\tau-\tau_{2}), \ \ p_1^{\prime}+p_2^{\prime}=1.
\end{eqnarray}
We apply this distribution of delays on three architectures:
(i) random; (ii) identical internal and external delays, model A (see Fig. \ref{fig:2}); and (iii) different internal  and randomly, equally-distributed  external delays, model B. 
Besides representing distinct phenomenological structures, these are motivated from the connectome. Its simplest decomposition on a left and a right hemisphere identifies the  peaks in 
$h(\tau)$
as 
internal and external
links, Fig. \ref{fig:1} (b-c), leading to model A. Other more complex divisions of the brain network could possibly identify patterns 
of 
 the other 
 architectures, or their combination. 

To preserve the distribution $h(\tau)$, which is over the links, the division of the nodes, $p_{1,2}$,  for model A  satisfies 
$[p_1^{\prime 2} + (1-p_1^{\prime})^2]/[2p_1^{\prime} (1-p_1^{\prime})]=p_1/(1-p_1)$, which gives
$p_{1,2}=\operatorname{Re}\{[1\mp\sqrt{\pm(1-2p_{1,2}^{\prime})}]/2\}$,
whereas for model B,
$p_{1,2}=p_{1,2}^{\prime}$.
The global order parameters hence read 
\begin{eqnarray}
 \label{eqn:z2}
  z(t) &=& p_1 z^{I}  + p_2 z^{II}.
 \end{eqnarray}
Superscripts correspond to the particular populations of models A or B, whereas because of the spatial homogeneity for the non-structured network,  $z^{I,II}=z$ can represent any proportion of nodes.
Similarly, the homogeneity of the internal links delays of subpopulations implies
\begin{eqnarray}
 \label{eqn:ksi2}
\xi_i^{I,II}(t) = \xi^{I,II}(t) = z^{I,II}(t-\tau_{int}) = z^{I,II}_{t-\tau_{int}},
 \end{eqnarray}
where $\tau_{int}$ are the internal delays of the populations.
Substituting Eqs.~(\ref{eqn:z2}, \ref{eqn:ksi2}) to Eq.~(\ref{eqn:KMdel}),  governing equations for all three topologies read
\begin{eqnarray}
 \label{eqn:KMdelMF1}
\dot{\theta}_i \ \  = & \omega_i \ \ - K [p_{1}  r_{t-\tau_1}   \sin(\theta_i - \psi_{t-\tau_1}) + \nonumber \\ 
 & \ \ \ \ \ \ p_{2} r_{t-\tau_2}  \sin(\theta_i - \psi_{t-\tau_2})]. \ \  \\
  \label{eqn:KMdelMF2}
 \dot{\theta}_i^{I,II} = & \omega_i^{I,II} -  K [p_{1,2} r_{t-\tau_1}^{I,II}  \sin(\theta_i^{I,II} - \psi_{t-\tau_1}^{I,II}) + \ \ \ \ \nonumber \\ 
 & \ \ \ \ \ p_{2,1} r_{t-\tau_2}^{II,I}   \sin(\theta_i^{I,II} - \psi_{t-\tau_2}^{II,I} )]. \ \  \\
  \label{eqn:KMdelMF3}
\dot{\theta}_i^{I,II} = \ & \omega_i^{I,II} -   K \{ p_{1,2}  r_{t-\tau_{1,2}}^{I,II}  \sin(\theta_i - \psi_{t-\tau_{1,2}}^{I,II}) + \ \ \ \ \nonumber \\  
p_{2,1}/2[r_{t-\tau_{1}}^{II,I}&\sin(\theta_i - \psi_{t-\tau_{1}}^{I,II})  + 
 r_{t-\tau_{2}}^{II,I}   \sin(\theta_i - \psi_{t-\tau_{2}}^{II,I})  ] \}. \ \  \ \ \ \
\end{eqnarray}

\section{Low-dimensional dynamics}
\label{sec:OA}

For infinitely large populations the dynamics of the system is described by PDFs for the phases of single oscillators, $\rho^{I,II}(\theta,\omega, t)$, which using continuum limit of Eqs.~(\ref{eqn:KMdelMF1}, \ref{eqn:KMdelMF2}, \ref{eqn:KMdelMF3}) evolve according to continuity equations
\begin{eqnarray}
\label{eqn:continuity}
 \frac{\partial\rho^{I,II}}{\partial t}=-\frac{\partial}{\partial \theta^{I,II}} (\dot{\theta}^{I,II}  \rho^{I,II} ).
\end{eqnarray}
Applying the OA ansatz \cite{Ott:08}, PDFs for the phases yield
 \begin{eqnarray}
\rho^{I,II}(\theta,\omega, t)=\frac{g(\omega)}{2\pi}\{1+\sum_{k=1}^{\infty} [\alpha^{I,II  k}(\omega, t)\mathrm{e}^{i k \theta}+c.c]\},  \nonumber
\label{eqn:OAansatz}
\end{eqnarray}
and consequently the global order parameters become
\begin{eqnarray}
\label{eqn:zInt}
z^{I,II}(t) &=&  \int_{-\infty}^{\infty}  \alpha^{I, II  \ast} ( \omega,t) g(\omega) d\omega. 
\end{eqnarray}
For a Lorentzian distribution
$g(\omega)=\gamma/\pi/[(\omega-\mu)^2+\gamma^2]$ 
with mean $\mu$ and scale  $\gamma$,
the populations' 
low dimensional dynamics \cite{omel2008chimera, Petkoski:12, Montbrio:11b, Ott:08, lee2009large, iatsenko2013stationary, barabash2014homogeneous} 
 become
\begin{widetext}
\begin{eqnarray}
\label{eqn:ld1}
\dot{z}  &=& (i \mu - \gamma) z- 
K/2[p_{1,2}(z^{2} \ z_{t-\tau_{1,2}}^{\ast} - z_{t-\tau_{1,2}}) +  
p_{2,1}(z^{2} z_{t-\tau_{2,1}}^{\ast} -z_{t-\tau_{2,1}})] . \ \\
\label{eqn:ld2}
\dot{z}^{I,II} &=& (i \mu - \gamma) z^{I,II} -
K/2 [p_{1,2}(z^{I,II \ 2} \ z_{t-\tau_{1}}^{I,II \ \ast} - z_{t-\tau_{1}}^{I,II} )+  
p_{2,1}(z^{I,II \ 2} z_{t-\tau_{2}}^{II,I \ \ast} -z_{t-\tau_{2}}^{II,I})] . \ \\
\label{eqn:ld3}
\dot{z}^{I,II} &=& (i \mu - \gamma) z^{I,II} - 
K/2 \{p_{1,2}(z^{I,II \ 2} \ z_{t-\tau_{1,2}}^{I,II \ \ast} -z_{t-\tau_{1,2}}^{I,II}) +
p_{2,1}/2[z^{I,II \ 2} (\ z_{t-\tau_{1}}^{II,I \ \ast} + \ z_{t-\tau_{2}}^{II,I \ \ast}) - \ z_{t-\tau_{1}}^{II,I} - \ z_{t-\tau_{2}}^{II,I} ] \}. \ \ \ 
\end{eqnarray}
\end{widetext}

\subsection{Critical couplings}
\label{subsec:Kc}
The incoherent state, $\{ z=0, \ \rho= 1/2 \pi\}$,  is a trivial solution to these systems, and due the non-negative inter-population contributions, Eqs.~(\ref{eqn:ld2}, \ref{eqn:ld3}), the possibility of only one incoherent populations is restricted. 
The lowest couplings for which the incoherence becomes unstable and synchronisation appears are determined from the purely imaginary eigenvalues of Jacobian matrices of the vector $[z_I, z_{II}]^T$ for structured, and of $z$ for random heterogeneity. 
For the latter these are solutions of 
\begin{eqnarray}
\label{eqn:Kc1}
\gamma + i (\beta - \mu) = K/2 [p_1 \mathrm{e}^{-i \beta \tau_1} +  p_2 \mathrm{e}^{-i \beta \tau_2}], \ 
\end{eqnarray}
whereas for models A and B respectively 
the following global conditions appear
\begin{eqnarray}
\label{eqn:Kc2}
&&[\gamma + i (\beta - \mu)]  [\gamma + i (\beta - \mu)- K/2 \  \mathrm{e}^{-i \beta \tau_1}] \nonumber  \\  
&& \   =  p_1p_2  K^2/4( \mathrm{e}^{-i 2 \beta \tau_2} -  \mathrm{e}^{-i 2 \beta \tau_1} ), \  \\
&&[\gamma + i (\beta - \mu)]  [\gamma + i(\beta - \mu) - K/2 (p_1\mathrm{e}^{-i \beta \tau_1} +p_2\mathrm{e}^{-i \beta \tau_2})] \   \nonumber  \\ 
&&\  = p_1p_2  K^2/16 ( \mathrm{e}^{-i \beta \tau_2} - \mathrm{e}^{-i  \beta \tau_1})^2. \ 
\label{eqn:Kc3}
\end{eqnarray}
It is worth noting that  the evolutions of $z^{I,II}$, Eqs.~(\ref{eqn:ld1}, \ref{eqn:ld2}, \ref{eqn:ld3}), depend only on the first Fourier harmonics of $\rho^{I,II}$, Eq.~(\ref{eqn:OAansatz}), and the same harmonic is the only one left in the linearised continuity equation (\ref{eqn:continuity}) \cite{yeung1999time}, hence, 
studying the dynamics of a small perturbation in the PDF of the phases $\rho^{I,II}(\theta,\omega, t)$ would yield the same conditions. 

\begin{figure}[t!]
\centering
\includegraphics{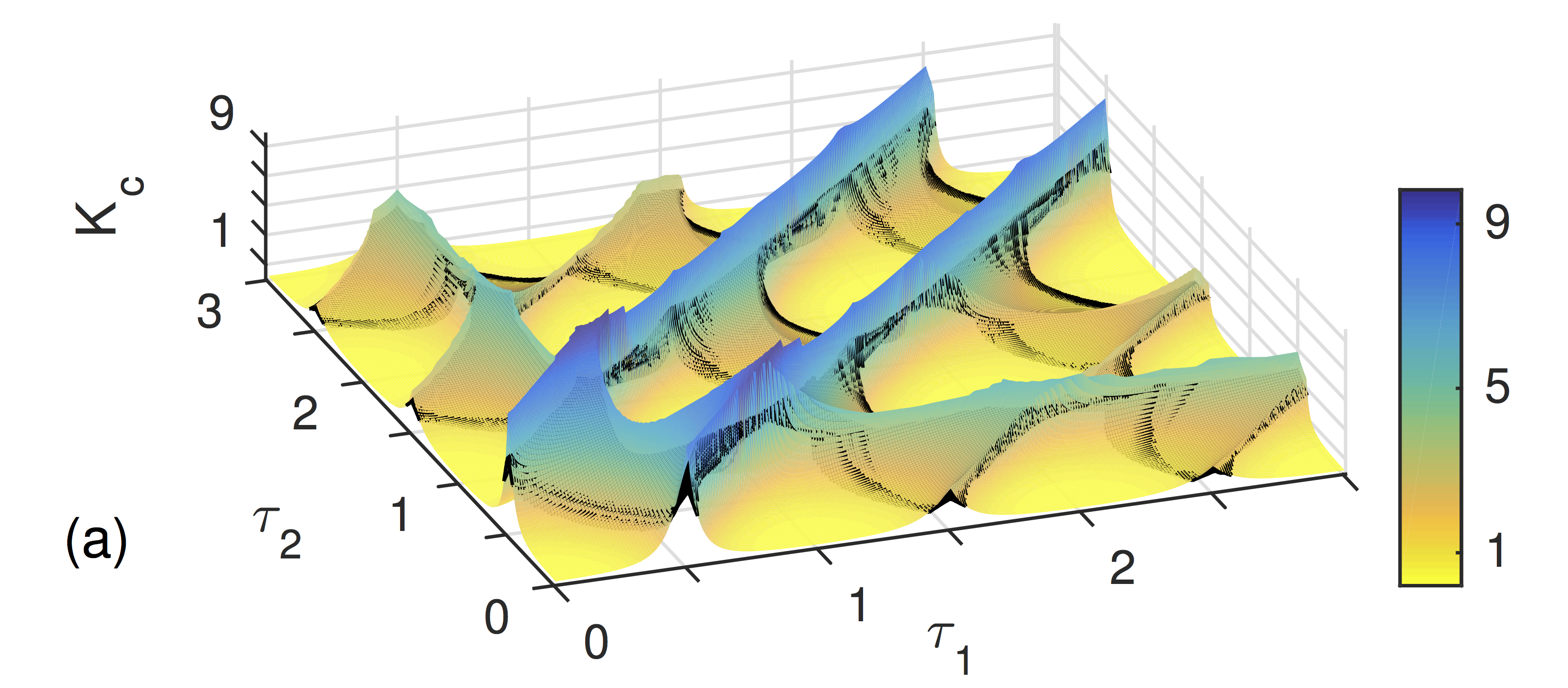}
\includegraphics{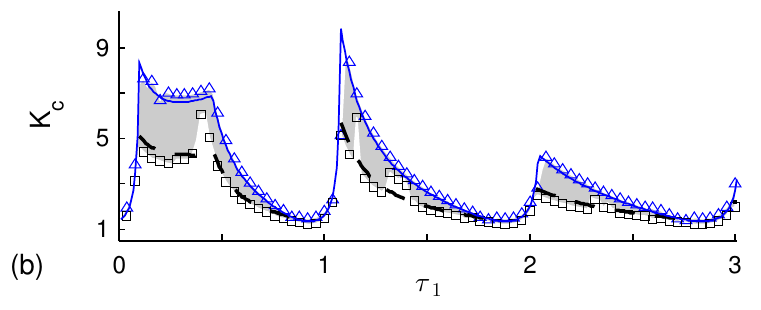} 
\includegraphics{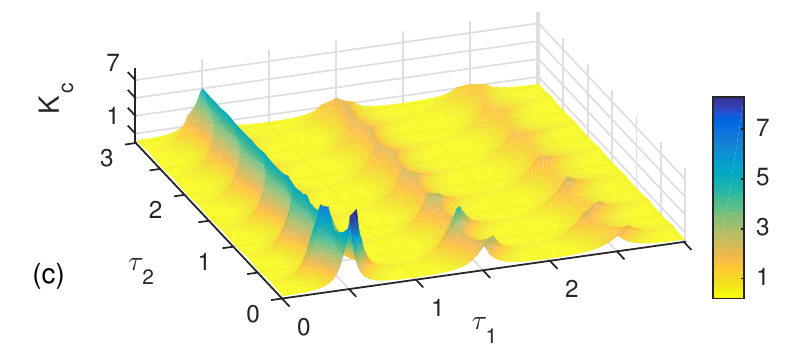} 
\includegraphics{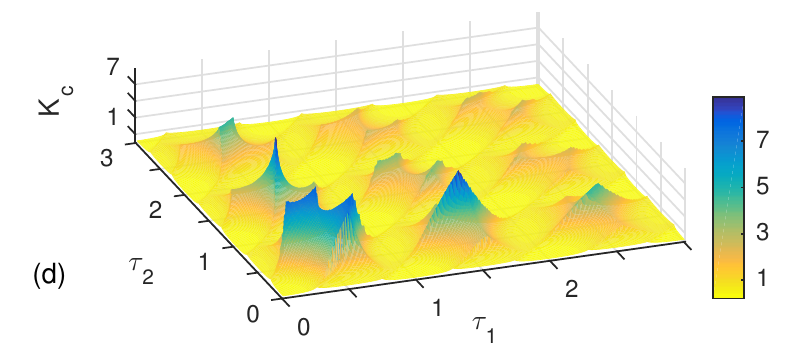} 
\caption{(color online) Critical couplings for incoherence for random bimodal-$\delta$ delays, Eq.~(\ref{eqn:Kc1}), (a, b), and for structured models A, Eq.~(\ref{eqn:Kc2}), (c) and B, Eq.~(\ref{eqn:Kc3}), (d). (a) Lowest couplings $K<K_c$ (black surface) for steady, Eqs.~(\ref{eqn:r2}), stable, Eq.~(\ref{eqn:multi}) solutions. 
(b) Intersection of plot (a) for $\tau_2=0.6$ (upper surface is thin blue and lower is dashed black line) and numerical results, Eq.~(\ref{eqn:KMdel}), for incoherence (blue trriangles),  coherence (black squares) and bistabillity (grey shaded areas). 
Parameters: $\mu=2\pi$, $\gamma=0.1$
and $p_1=0.5$.}
\label{fig:3}
\end{figure}
The critical couplings, Fig. \ref{fig:3}, show the crucial role of the delay's topology in shaping the synchronization landscape.
For the case with no structure in the couplings, the ridges of $K_c$  are highest at 
$|\tau_1 - \tau_2| =  n T/2$, 
where $T=2\pi/\mu$ is a mean period of the natural frequencies and n is a positive odd integer. 
These are followed by smaller peaks at $\tau_{1,2}= nT/2$, Fig. \ref{fig:3} (a, b).
%
%
For model A, the internal delays ($\tau_{int}=\tau_1$ in Fig. \ref{fig:3} (c)) are the main factor for preventing the synchronization 
 around $nT/2$ , same as for unimodal delays \cite{yeung1999time, lee2009large}.
On the other hand, the interpopulation influence is $T/2$ periodic, with largest $K_c$ around  $\tau_2=nT/4$.
This is due to the anti-phase arrangement of the synchronized populations (see Fig. \ref{fig:4}), which causes them to enhance coherence for $\tau_2$ around $nT/2$.
For  model B, the synchronizabillity  is more complex, but it is still  a combination of the former two: $K_c$ has $T$ periodic peaks at $\tau_{1,2}=n T/2$ due to the internal antiresonance of populations, and $T/2$ periodic at $|\tau_1 - \tau_2|=nT/4$, corresponding to the interpopulation influence, Fig. \ref{fig:3} (d). 
In all scenarios the peaks are dampened at consecutive periods.


Randomly  distributed delays imply spatial homogeneity, where each oscillator is forced by a mean field from oscillators linked with delay $\tau_1$, and from those with $\tau_2$, Eq.~(\ref{eqn:KMdelMF1}). These local order parameters, Eq.~(\ref{eqn:ksi}), are at distance $\Omega(\tau_2-\tau_1)$, where $\Omega$ is the frequency of synchronization.
For a steady, $\dot{r}=0$, travelling wave synchronization \cite{petkoski2013mean, iatsenko2013stationary}, 
the mean field, $z=r \mathrm{e}^{\Omega t}$, has parameters
 \begin{eqnarray}
\Omega &=& \mu - K/2(r^2+1) (p_1 \sin\Omega \tau_1  + p_2 \sin \Omega \tau_2 ), \nonumber \\ 
r &=& \sqrt{ 1- 2 \gamma/[K (p_1 \cos\Omega \tau_1  + p_2 \cos \Omega \tau_2 ) ]}.
\label{eqn:r2}
\end{eqnarray}
Their stability is obtained by introducing a small perturbation 
$\delta z =(\mathrm{e}^{(\lambda +i \beta) t} + \mathrm{e}^{(\lambda -i \beta) t}) \mathrm{e}^{i \Omega t}$,
and looking for solutions of  Eq.~(\ref{eqn:ld1}) with non-negative $\lambda$   \cite{lee2009large}. 
This  yields 
 \begin{eqnarray}
\label{eqn:multi}
\gamma-i(\Omega-\mu-\beta) +  K/2 (p_1A_1+p_2A_2 )=0, \ \ \
\end{eqnarray}
where $A_{1,2}=2r^2\mathrm{e}^{-i \Omega\tau_{1,2}}-\mathrm{e}^{-i(\beta-\Omega)\tau_{1,2}}-r^2\mathrm{e}^{-i (\beta+\Omega)\tau_{1,2}}$, 
and if solutions exist, then 
$r \mathrm{e}^{\Omega t}$
 is not stable. Hence,  the lower bounds of synchronization, the black surface in Fig. \ref{fig:3} (a), are determined by exploring stable solutions for $K<K_c$, and they follow the same resonant patterns as the critical couplings.
Numerical results in Fig. \ref{fig:3} (b) confirm the regions where the population either synchronises, or becomes incoherent, for any of the many initial states in the range  $r(t)|_{t\in[-\max(\tau),  0]}=[0, 1]$ 
that we checked. 
However, time delays imply infinitely many possible initial states and it is very probable that some of the solutions have extremely small  basins of attractions, as was also reported in  \cite{choi2000synchronization, lee2009large}, and are difficult to be numerically recovered.

Bistabillity, where synchronizabillity depends on the initial history, was recovered in between the critical surfaces. 
The integration was performed on both, the full system with  $N=1000$ oscillators, Eq.~(\ref{eqn:KMdel}) and the low-dimensional dynamics, Eq.~(\ref{eqn:ld1}). 
Interestingly, beside multiple stable solutions of Eq.~(\ref{eqn:r2}), the same procedure unveils non-steady synchronization, $\dot{r}\neq 0$,  for initial states close to the incoherence, in regions around $\tau=[nT/2, (n+1) T/2]$ that decrease in size for increasing $n$ \cite{petkoski:inprep}.
These are  consequence of the bimodallity of delays, since they are not reported for non-identical oscillators with homogeneous \cite{yeung1999time, choi2000synchronization} or with unimodal heterogeneous delays \cite{lee2009large}.

\subsection{In and anti-phase clustering}
\label{subsec:pi}

\begin{figure}[t!]
\centering
\includegraphics[width=0.4\textwidth]{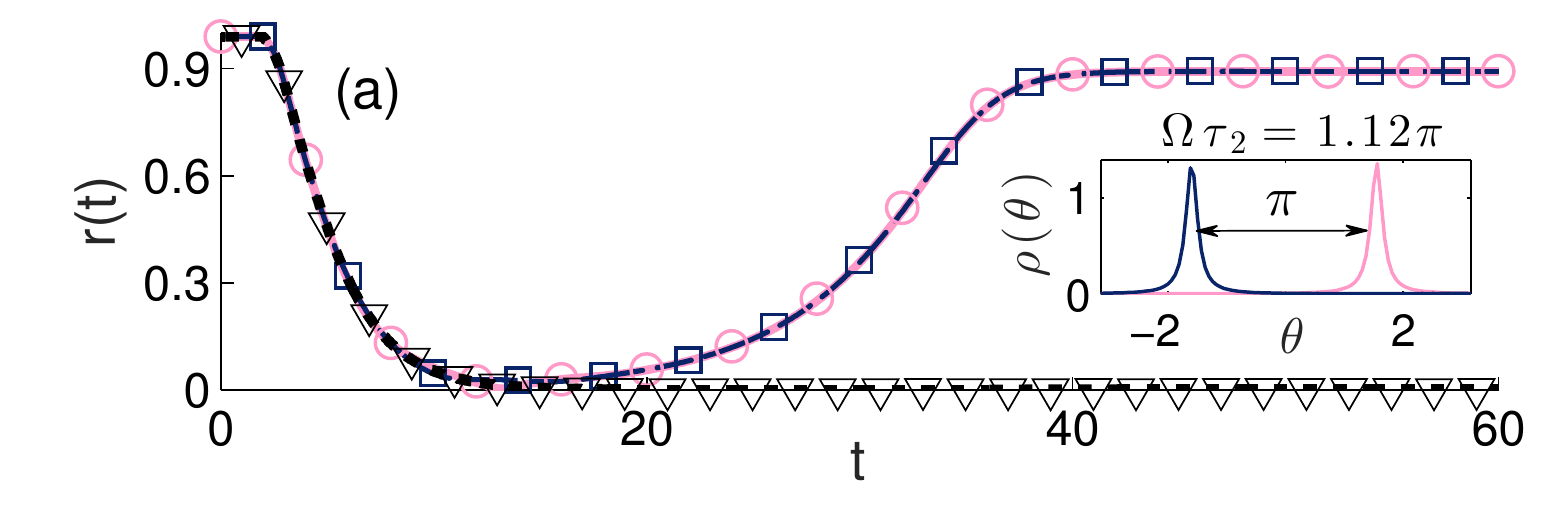}  
\includegraphics{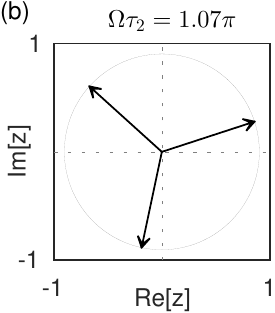} \hspace{2em} 
\includegraphics{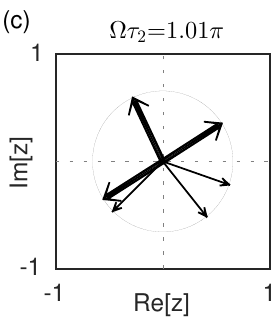} 
\caption{Anti-phase clusters for model A with (a) 2, (b) 3 and (c) 9  equal populations. 
(a, b)	Blue and red are for the first and second, and grey is for the overall population. 
(a) Magnitude of order parameters, 
theoretical (dashed blue, thick red and dotted grey), Eq.~(\ref{eqn:ld2})  and  numerical results (blue squares, red circles and grey triangles), Eq.~(\ref{eqn:KMdel}) for $N=100000$ oscillators.
Inset: PDF of the phases.
(b, c) Order parameters $z_{m}$. 
(c) Each of the three thicker arrows accounts for two identical mean-fields. 
 Parameters:  $K=2$, 
 $\mu=2\pi$, $\gamma=0.1$,
 (a) $\tau=[0.3, 0.7]$, (b, c) $\tau=[0.15, 0.55]$. 
}
\label{fig:4}
\end{figure}

In model A, if equally divided, identical $r$ and $\Omega$ can be assumed for each populations.
 Stationarity conditions then read
\begin{eqnarray}
z^{I}= r\mathrm{e}^{i\Omega t}, \ z^{II} = r\mathrm{e}^{i(\Omega t + \phi)},\ \dot{r}= 0, \ \Delta\dot{\phi}_{21}=\dot{\phi}= 0, \ \ \ \
\label{eqn:stat}
\end{eqnarray} 
and substituting them into Eq.~(\ref{eqn:ld2}), yields
$$\dot{\phi} = -K/2(r^2 + 1) \cos\Omega \tau_2   \sin\phi.$$
This implies $\phi=0\pm\pi$, with stable zero phase shift for $\Omega\tau_2\in(-\pi/2, \pi/2)$ and  $\phi=\pi$ stable otherwise.  
Thus, by increasing the  delay between the populations, 
they rearrange from  in- to  anti-phase, Fig. \ref{fig:4} (a). 
Notably, non-steady states occur for certain low coherence initial states, around the same parameter's space as for the random case \cite{petkoski:inprep}.

The same clustering phenomenon persists for more than two equal populations: their order parameters are identical for 
$\Omega \tau_2 $ in the right half-plane,
 or they arrange to cancel each other otherwise.
 In this case however the PDF of the delays Eq.~(\ref{eqn:h})  changes and since the system is symmetric the distribution of the links follows the division of the network nodes.
 Thereupon,  
$$ h(\tau)={1}/{M}\delta(\tau-\tau_1)+{(M-1)}/{M}\delta(\tau-\tau_2), $$
and the model no longer corresponds to a simple spatial rearranging of the same global distribution of time-delays. 
Continuing Eq.~(\ref{eqn:ld2}) for $M$ populations with $p_m=1/M, \ \forall m\in[1..M]$,  mean-phases $\phi_m$ evolve as
\begin{eqnarray}
\dot{\phi}_m=\mu-\frac{Kr(r^2+1)}{2M} [\sin\Omega\tau_1+\sum_{j\neq m} \sin(\Omega\tau_2+\Delta\phi_{mj})]. \ \ \ \ \
\label{eqn:phiAm} 
\end{eqnarray} 
Therefore the stationarity of  $r$ and $\Delta\dot{\phi}_{mn}$  for all pairs of populations $(m,  n)$ implies
\begin{eqnarray}
 \sum_{j} \mathrm{e}^{i\Delta\phi_{mj}} =\sum_{j} \mathrm{e}^{i\Delta\phi_{nj}}. \nonumber
 \label{eqn:anticond}
\end{eqnarray} 
This  is  satisfied either if all complex order parameters are aligned, or if they 
cancel each other in the stationary state. 
For the latter, an equidistant arrangement is the exclusive pattern for 2 and 3 populations
with mean phase distances of $\pi$ and $2\pi/3$ respectively,
 Fig. \ref{fig:4} (a, b). 
 For $M>3$ this is no longer unique and there are infinite possible arrangements. 
 In some of these it is possible identical order parameters to occur for some of the populations, while the sum of all of them is still 0, Fig. \ref{fig:4} (c).
As for the mean-field parameters, their steady state values are given by 
 \begin{eqnarray}
\label{eqn:r1}
\Omega &=& \mu-  M \frac{K}{2} (r^2+1) [ \sin\Omega \tau_1  +  \sum_{j\neq m}  \sin (\Omega \tau_2 + \Delta\phi_{mj}) ], \nonumber  \\ 
r &=& \sqrt{ 1- \frac{2 M \gamma} { K [ \cos  \Omega \tau_1  +  \sum_{j\neq m}  \cos ( \Omega \tau_2 + \Delta\phi_{mj} ) ] } }. \ \ 
\end{eqnarray}
For in-phase arrangement, $ \Delta\phi_{mj}=0, \ \forall \ m, j \  \in[1..M] $, and the above trigonometric sums become
$(M-1)  \sin \Omega \tau_2$  and  
$(M-1)  \cos \Omega \tau_2$, respectively, 
whilst for the anti-phase state they yield $ - \sin \Omega \tau_2$  and  
$|\cos \Omega \tau_2|$.
Consequently, increasing the number of populations decreases the level of coherence, as shown in Fig. \ref{fig:4} (b-c) for 3 and for 9 populations.


\subsection{Time-varying synchronization}
\label{subsec:TV}

\begin{figure}[t!] 
\centering
\includegraphics[width=0.45\textwidth]{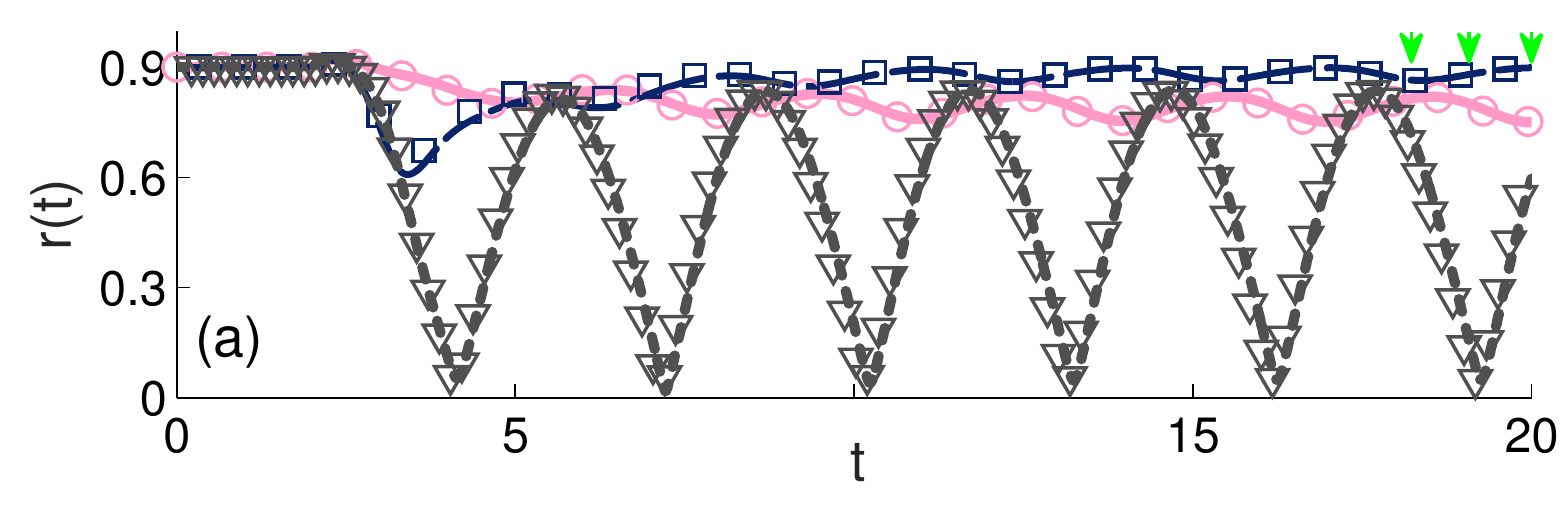} 
\includegraphics[width=0.45\textwidth]{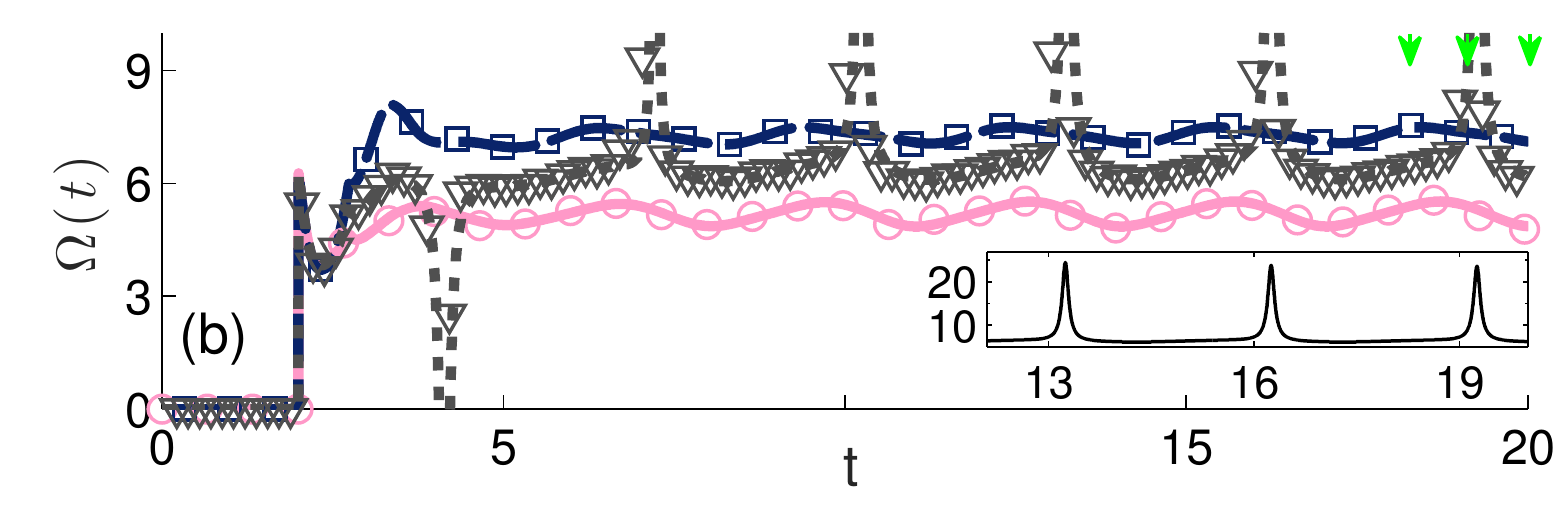} 
\includegraphics[width=0.47\textwidth]{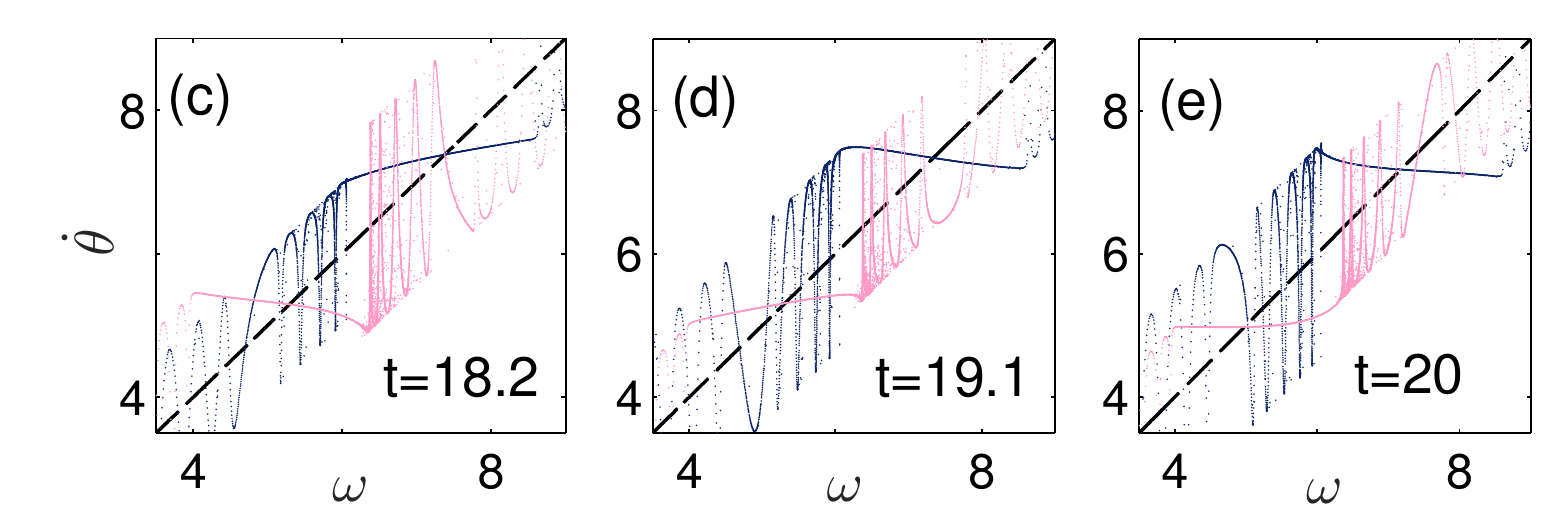} 
\caption{(color online) (a) Magnitudes of non-steady global order parameters, (b) mean field frequencies, and (c - e)  adjusted frequencies for model B. (Green) arrows in (a, b) indicate the time points $t=[18.2, 19.1, 20]$ for the results in plots (c - e). Blue, red and grey correspond to the first, second, and the overall population.  (a, b) Theoretical (dashed blue, thick red and dotted grey), Eq.~(\ref{eqn:ld3}), and numerical results (blue squares, red circles and grey triangles), Eq.~(\ref{eqn:KMdel}). Parameters:  $K=3$, $p_1=0.5$,  $\mu=2\pi$, $\gamma=0.1$,
$\tau=[0.23, 0.74]$, $N=100000$.} 
\label{fig:5}
\end{figure}
For model B with equal populations, assuming that they settle to same  $r$ and  $\Omega$,  Eqs.~(\ref{eqn:ld3}, \ref{eqn:stat})  give
\begin{eqnarray}
\hspace{-0.3em}
\label{eqn:phidot2}
\dot{\phi}=-\frac{K}{4}(r^2+1)(\cos\Omega \tau_1+\cos{\Omega \tau_2})(\sin\phi+\tan{\frac{\Omega\Delta\tau}{2}}), \ \ \ \ \  \\
\hspace{-0.3em}
\Delta\dot{r}=\frac{Kr}{4}(r^2-1)(\sin\Omega\tau_1+\sin{\Omega\tau_2})(\sin\phi+\tan{\frac{\Omega\Delta\tau}{2}}). \ \ \ \ \  
\label{eqn:rdot2}
\end{eqnarray}
Hence,   $|\Omega \Delta \tau|  \in (-\pi/2, \pi/2)$ is a necessary condition for this stationarity and the synchronized clusters are at distance $\phi=-\arcsin\tan(\Omega\Delta\tau/2)$.
However, if $|\Omega \Delta \tau|  \notin (-\pi/2, \pi/2)$, Eqs.~(\ref{eqn:phidot2}, \ref{eqn:rdot2}) cannot be zero and these instabilities continuously  persist, implying non-stationary synchronization,  
 Fig. \ref{fig:5} (a). This
 is characterized with fast spikes of the overall mean frequency,  Fig. \ref{fig:5} (b), and continuous  rearrangement  of oscillators, so that some of them are always entrained  with the mean field of the other population, as can be  seen from the adjusted  versus the natural frequencies of the individual oscillators, $\dot{\theta}(\omega)$, captured at different moments in Fig. \ref{fig:5} (c-e).  
Contrary to the previous scenarios, the non-steady states here can appear for all initial conditions and for a wider parameter space \cite{petkoski:inprep}.

\subsection{Realistic brain delays and EEG frequencies}
\label{subsec:TV}

\begin{figure}[tb!] 
\centering
\includegraphics{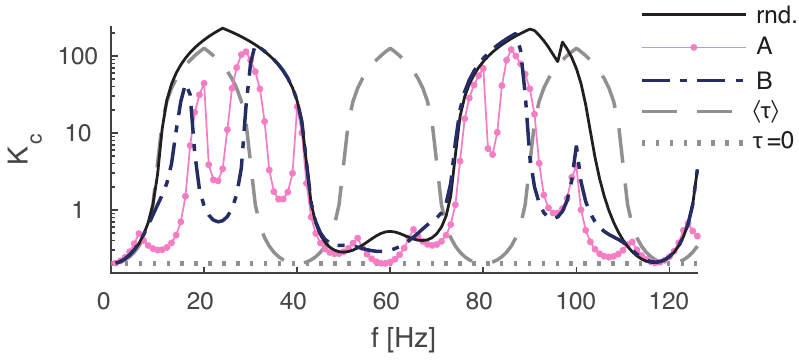} 
\caption{ Critical couplings for synchronization at EEG frequencies for all-to-all network with identical coupling strengths and different bimodal-$\delta$ spatial distributions of realistic brain delays, $\tau=[$18ms, 42ms$]$, $p_1=0.7$. 
} 
\label{fig:6}
\end{figure}

If we set identical propagation velocity from within the physiological range for brain signals \cite{nunez2006electric}, e.g.  at 2 m$/$s, then for the human brain tract lengths data shown in Fig. \ref{fig:1},  the time-delays due to tracts have peaks around 18 ms and 42 ms with proportion $p_1=0.7$ for the the bi-modal $\delta$  approximation  Eq.~(\ref{eqn:h}). 
Taking these values for networks with Lorentizan natural frequencies with spread $\gamma=0.1$ rad$/$s as in the earlier examples, but with means $\mu$ at realistic EEG frequency range, using Eqs.~(\ref{eqn:Kc1}, \ref{eqn:Kc2}, \ref{eqn:Kc3}) we calculate  the regions of synchronizabillity for  the three discussed delay-imposed networks (random, model A and model B), Fig. \ref{fig:6}.  
For comparison, the critical coupling for an identical delay $\langle \tau \rangle$ corresponding to the mean of all delays $\langle h(\tau)\rangle$ for these parameters, and for absent delays, $ \tau   = 0$, is also shown.

The positions of the peaks of the critical couplings $K_c$ relatively to the period $T$ are as discussed  in Sec. \ref{subsec:Kc}, with the delays now being fixed, instead of the natural frequencies. 
We see that  $K_c$ can differ by 3 orders of magnitude depending on the frequency and the network architecture, with bimodal scenarios  showing distant patterns from the case with a single delay, while each structured case is also specific.  

Note that for fully realistic description of the brain synchronizabillity, the heterogeneous connectivity weights, Fig. \ref{fig:1} (a), which  imply a  complex network  \cite{Rodrigues2015}, would also need to be taken into account. 

\section{Summary}
\label{sec:Sum}

Transforming many time delays into spatial patterns within the couplings' space-time structure provides a novel concept 
for a better understanding of large-scale network dynamics. 
Together with the various dynamical regimes discussed earlier, these findings unveil the critical importance of
spatial heterogeneity of the
 time-delays  in the coupling matrix.
 Unlike populations defined by coupling strengths or natural frequencies \cite{Montbrio:04}, 
the structure 
here
stems solely from the link-delays, and introduces non-trivial spatiotemporal dynamics compared to 
homogeneous  \cite{yeung1999time, choi2000synchronization, barabash2014homogeneous}, or random unimodal heterogeneous delays \cite{lee2009large}.
Future work should extend these results for other coupling functions of the phase reduced model, e.g. similar to those in \cite{Roy2011}, and for realistic neurons.

We have here provided a theoretical framework, which allows treating the space-time structure of couplings as a whole with regard to its effects upon network synchronization. 
 A relevant real-world example is found in clinical neuroscience, where the reshaping of the 
 time delays is common in neurodegenerative diseases such as multiple sclerosis \cite{lucchinetti2000heterogeneity}, but also known to be critical in aging \cite{sullivan2010quantitative} and neuroplasticity \cite{pascual2005plastic}.

Finally, anti-phase spatio-temporal brain patterns as a paradigm \cite{Li2011}, analogous to those observed for model A as a first approximation of the connectome, have been observed and modelled across different frequency bands, and imaging \cite{ Deco:09} and electrophysiological data \cite{Szucs2009}.
Similarly, many recent models for the pair-wise coherence in connectome-based networks of phase oscillators \cite{ghosh2008noise, Cabral2014}
that try to reproduce the patterns of coherence and incoherence observed in resting brain, 
would be simplified by several orders of magnitude by applying our reduction.  This becomes even more important for finer brain parcellations, where the numerical analysis of the full-delayed system is tremendously computational extensive.

\section*{Acknowledgments}
The research reported herein was supported by the Brain Network Recovery Group through the James S. McDonnell Foundation, the Aix-Marseille Universit\'{e} foundation A$\ast$Midex - CoordAge project (ANR-11-IDEX-0001-02); and funding from the European Union Seventh Framework Programme: FP7 Human Brain Project (grant no. 60402).

\bibliography{distdelKM_refs}

\end{document}